\documentclass[12pt]{article}
\usepackage{amsmath}
\usepackage{amssymb}
\usepackage{array}
\usepackage{geometry}
\usepackage{graphicx}
\usepackage{enumerate}
\numberwithin{equation}{section}

\begin{document}
\begin{center}
{\bf \large SOLAR NEUTRINO RECORDS:\\ 
GAUSS OR NON-GAUSS IS THE QUESTION}\\
\bigskip
{\bf A. Haubold$^1$, H. J. Haubold$^{2,3}$ and D. Kumar$^3$  }\\
\small{$^1$ Department of Computer Science, \\
Columbia University, New York, NY 10027, USA\\
Email: ah297@columbia.edu\\
$^2$Office for Outer Space Affairs, United Nations, \\
Vienna International Centre, P.O. Box 500, A-1400 Vienna, Austria\\
Email: hans.haubold@unvienna.org\\
$^3$Centre for Mathematical Sciences Pala Campus\\
Arunapuram P.O., Palai, Kerala  686 574, India\\
Email: dilipkumar.cms@gmail.com, Website: www.cmsintl.org
}
\end{center}
\begin{center}
{\bf Abstract} \\
\end{center}

{ \small
This article discusses the possible variation of the solar neutrino flux over time in the records of Super-Kamiokande-I and the relation to non-equilibrium statistical mechanics the and entropic pathway model. The scaling behavior of the Super-Kamiokande-I time series is investigated utilizing Standard Deviation Analysis and Diffusion Entropy Analysis. The data set exhibit scaling behavior and may follow L\'{e}vy-type probability distribution function.\\


\noindent{\bf 1. Introduction}\\

In recent years, physicists and mathematicians have been discussing the possible signature of variation of the solar neutrino flux over time in the records of solar neutrino experiments and the relation of such a signature, if discovered, to non-equilibrium statistical mechanics as developed by Tsallis \cite{1,10} and to the entropic pathway as introduced by Mathai \cite{4}. Spatio-temporal analysis of solar activity and possible correlation to the variation of the solar neutrino flux, as reported in the literature, have also been debated \cite{11}. Such discussions were central to the UN/ESA/NASA/JAXA workshop for the preparation of the International Heliophysical Year 2007, Abu Dhabi and Al-Ain, United Arab Emirates, 20-23 November 2005 \cite{9}. This article reports some progress in the discussions made since then.\\

\noindent{\bf 2. Time Series: Variance Versus Probability Density Function}\\

Time series are recording the evolution of a parameter over time and commonly exhibit stochastic fluctuations of the recorded parameter if the underlying process is random. Traditional methods for the study of scaling exponents of such time series like standard deviation analysis and spectral analysis (Fourier, wavelet) are based upon the analysis of the variance of the time series \cite{6,7}.

The statistical variability of the average solar neutrino flux is interpreted as noise (see Fig.1). The solar neutrino flux fluctuations are thought to contain no useful information and are consequently smoothed and show a constant average in the long time. According to the central limit theorem, the statistics of the fluctuations in such a time series, emanating from complex systems such as the gravitationally stabilized solar fusion reactor, are assumed to be Gaussian. The fact that they may not be, due to the fact of being generated by an open system, remains unexplained. However, suspected non-Gaussian behavior prompted studies of possible time variation of solar neutrino records as a problem belonging to non-equilibrium statistical mechanics wherein statistical fluctuations can provide useful information about, for example, the reaction and diffusion properties of complex phenomena. An example would be the fluctuation-dissipation theorem, in which the response of a complex system to a perturbation is determined by the complex system's unperturbed autocorrelation function. The variation of the solar neutrino flux would be indicative of the Sun's complex internal dynamics as evidenced by changes in the number of detected solar neutrinos. Such studies maintain that the variations in the solar neutrino flux detected on Earth are not noise, but contain information about the source of variability, particularly the variations in the gravitationally stabilized solar fusion reactor, beyond neutrino oscillations \cite{8}.

Different from standard deviation analysis (SDA), Scafetta's method of diffusion entropy analysis (DEA) evaluates, under certain conditions, the scaling exponent of the probability density function (pdf) through the Shannon entropy (or Tsallis entropy, or Mathai's pathway) of the diffusion process generated by the fluctuations exhibited by the time series \cite{7}, such as shown in Fig.1. The application of DEA shows that the fluctuations in the solar neutrino record may have inverse power-law statistical distributions. Specifically, if $x$ is the solar neutrino flux, the distribution of the values of neutrino fluxes is an inverse power law $P(x) \sim \frac{C}{x^\alpha}$, where $C$ is a normalization constant. The parameter $\alpha$ is the inverse power-law index. The scaling of the solar neutrino record was tested by randomly changing the order of the data points. If the time series were internally correlated, the resulting distribution would have changed from the original, but that could not be observed. The invariance of the distribution under shuffling indicates that the statistics of the time series in non-Poisson and renewal, meaning that with the generation of each new event, the process is renewed \cite{12}.\\

\noindent{\bf 3. Diffusion: Gauss Versus L\'{e}vy Statistics}\\

The statistics of the solar neutrino record is described by a non-Gaussian distribution. The behavior of such limit distributions requires a generalization of the central limit theorem to the case in which the second moment of the variate diverges. Such processes were studied by L\'{e}vy. The solar neutrino record is describable by such a L\'{e}vy distribution.

The diffusion process, as utilized in Scafetta's DEA, is modeled as a continuous-time random walk where a random walker jumps instantaneously from one location in space to another following a waiting period on a location whose duration is drawn from a pdf of waiting times \cite{5,12}. Depending on the rules of walking in space and jumping over time, pdf's of the diffusion process show Gaussian behavior or can be represented by L\'{e}vy-type flights or L\'{e}vy-type walks \cite{3}. Such pdf's can be derived from Mathai's entropic pathway. In Mathai's pathway model, a specific parameter allows to proceed from probability densities of a generalized type-1 beta model to a generalized type-2 beta model to a generalized gamma model when the variable is restricted to be positive \cite{4}. See Figures 1 and 2.
\begin{center}
 \resizebox{4.5cm}{!}{\includegraphics{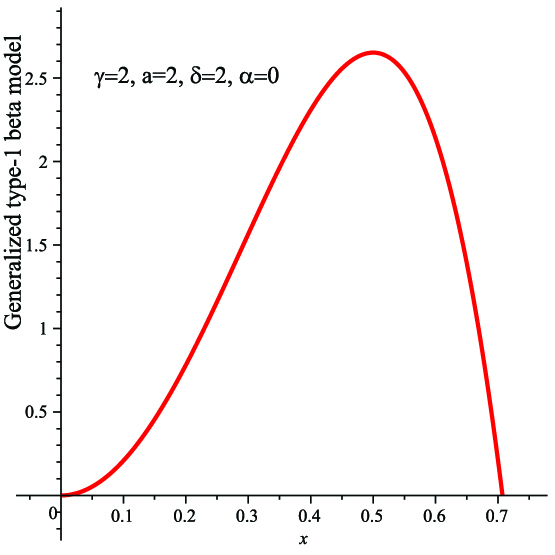}}(a)
 \resizebox{4.5cm}{!}{\includegraphics{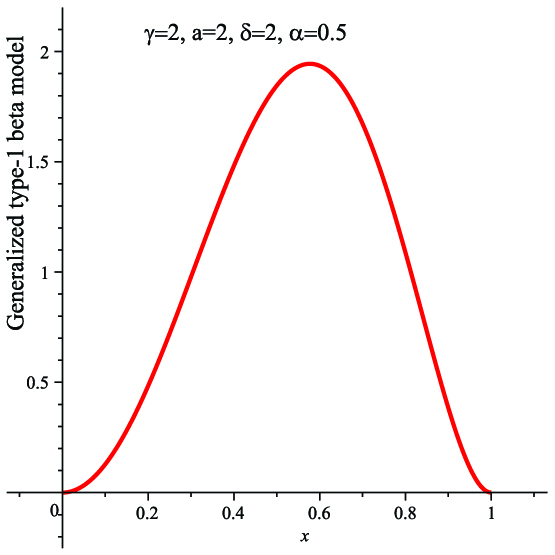}}(b)
 \resizebox{4.5cm}{!}{\includegraphics{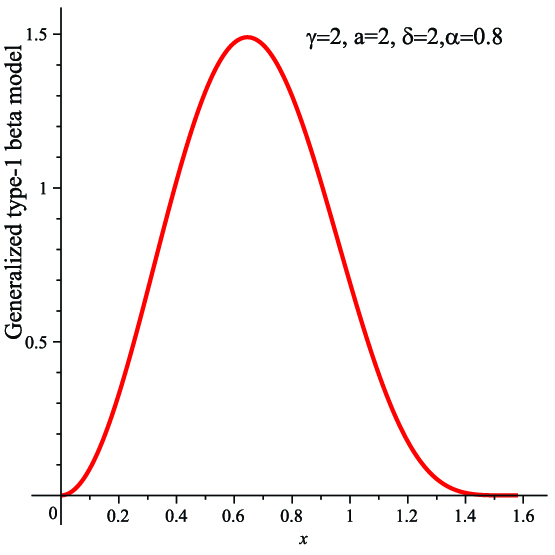}}(c)
\end{center}
The Figure 1 (a), (b) and (c) show the behaviour of the generalized type-1 beta form of the pathway model given by the function $f_1(x)=c_1x^{\gamma}[1-a(1-\alpha)x^\delta]^{\frac{1}{1-\alpha}}$ for the values $\gamma=2, a=2, \delta=2$ and for $\alpha=0, \alpha=0.5$ and $\alpha=0.8$ respectively. $c_1$ is the normalizing constant. This is a model with the right tail cut-off when $\alpha<1$.
\begin{center}
   \resizebox{7cm}{!}{\includegraphics{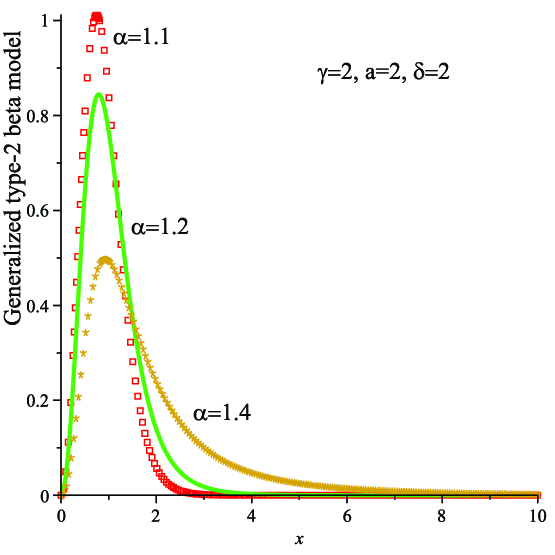}}(a)
    \resizebox{7cm}{!}{\includegraphics{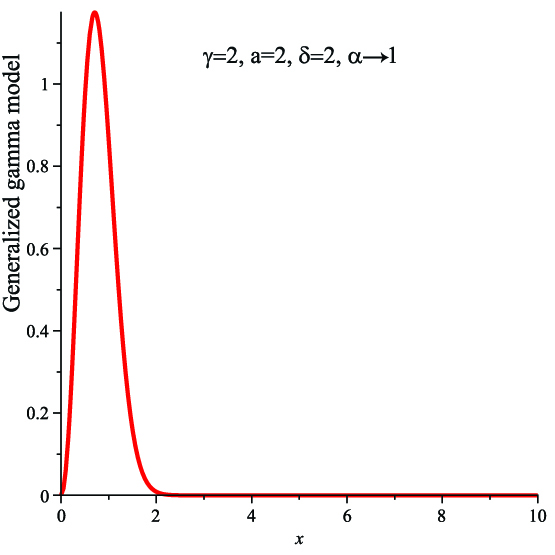}}(b)
 \end{center}
The Figure 2 (a) shows the behaviour of the generalized type-2 beta form of the pathway model given by the function $f_2(x)=c_2x^{\gamma}[1+a(\alpha-1)x^\delta]^{-\frac{1}{\alpha-1}}$ for the same values $\gamma=2, a=2, \delta=2$ and for $\alpha=1.1, \alpha=1.2$ and $\alpha=1.4$, respectively. Here $c_2$ is the normalizing constant. Superstatistics of Beck and Cohen \cite{1} belongs to this case. (b) shows the behavior of the generalized gamma form of the pathway model or the Gaussian model given by the density function $f_3(x)=c_3x^{\gamma}{\rm e}^{-a x^\delta}$ for the values $\gamma=2, a=2, \delta=2$.  As $\alpha \rightarrow 1$ the type-1 and type-2 beta models  $f_1(x)$ and $f_2(x)$ reduce to Gaussian form given by $f_3(x)$.
Taking into account correlations (memory), deviation from Gaussian behavior can be described by application of fractional derivatives in space and time. The diffusion process can be extended to a reaction-diffusion process if the species are not only mobile in space and time but are also allowed to react with each other \cite{4}. In this case, diffusion, i.e. random walk of reactants, is the mechanism that allows for reaction, in more general terms described by fractional reaction-diffusion equations. Closed-form representations of such equations, reflecting non-Gaussian behavior, are achieved by utilizing Mittag-Leffler functions and H-functions \cite{2,4}.\\

\begin{center}
 \resizebox{15cm}{!}{\includegraphics{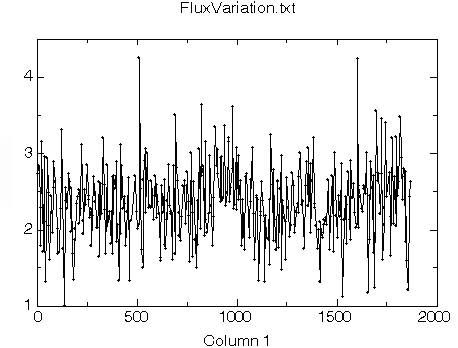}}
\end{center}
Figure 3 shows the horizontal axis with a time period $(1997-2002)$ from the beginning of the data taking and the vertical axis which is the measured neutrino flux in units of $10^6 cm ^{-2} s ^{-1}$. The $\frac{1}{R^2}$ correction is included in the measured solar neutrino flux of 5-day-long samples from Super-Kamiokande-I \cite{13}. It is commonly assumed that this record of data does neither contain a trend nor a periodicity. The average value of the record confirms the expected average solar neutrino flux based on a standard solar model.
\begin{center}
 \resizebox{15cm}{!}{\includegraphics{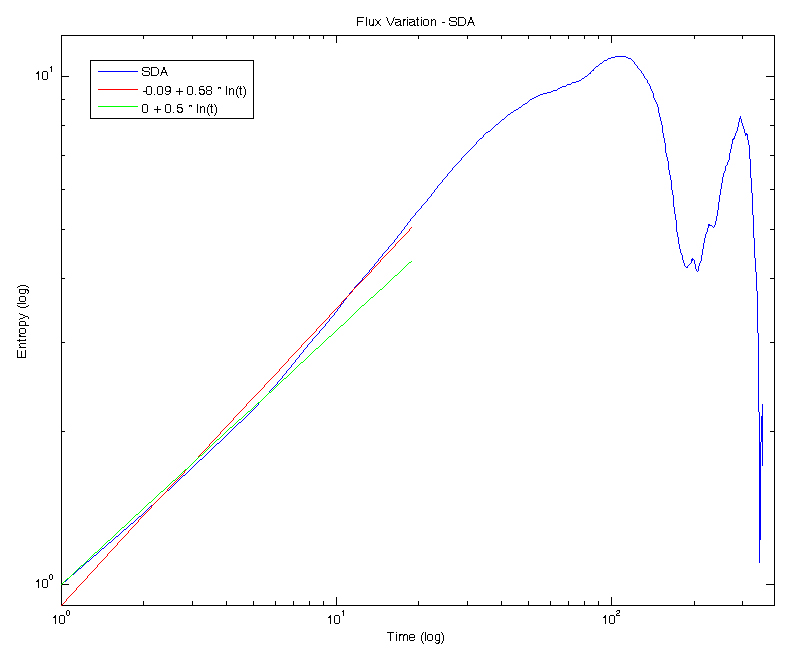}}
\end{center}
The Figure 4 shows the results of standard deviation analysis (SDA). In the SDA (blue line) one examines the scaling properties of the second moment of the diffusion process generated by the time series as shown in Fig. 1. The standard deviation $D(t)$ of the variable x is \[D(t) = (<x^2,t > - < x,t>^2)^\frac{1}{2}\sim  t^H.\] The Hurst exponent H is interpreted as the scaling exponent being evaluated from the gradient of the fitted straight line in the log-log plot of $D(t)$ over diffusion time $t$.
In the Figure, the Hurst exponent of $H=0.58$ for the red straight line does indicate a deviation from Gauss. Also shown in green is the line for $H=0.5$.

\begin{center}
 \resizebox{15cm}{!}{\includegraphics{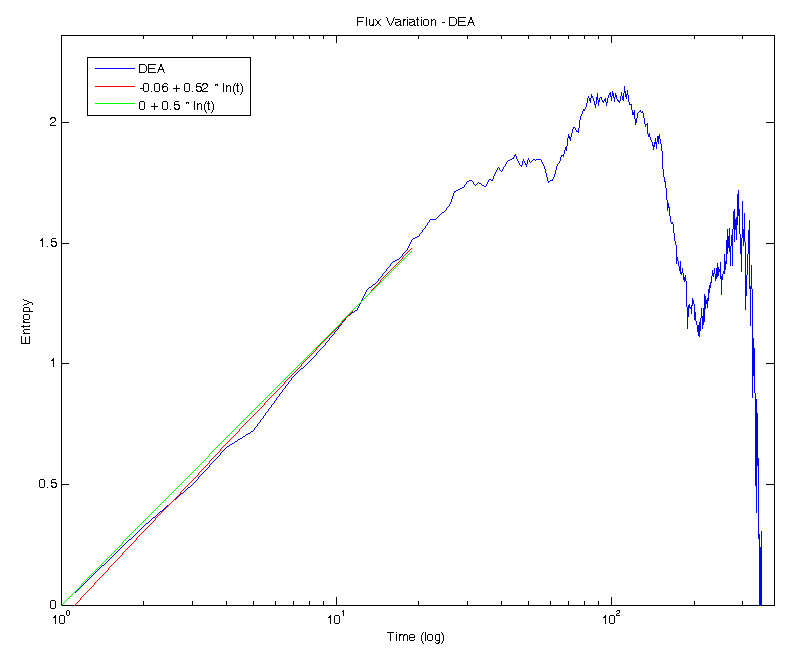}}
\end{center}
The Figure 5 shows the results of diffusion entropy analysis (DAS). The DEA determines the scaling exponent delta evaluated through the Shannon entropy $S(t)$ of the diffusion process generated by the variation of the time series in Fig. 1. The probability distribution function $p(x, t)$ is determined by means of the sub-trajectories \[X_n(t) = {\displaystyle \sum_{i=0}^l }\xi_{i+n} , n = 0, 1,\ldots. \] If the scaling condition \[p(x,t) = (\frac{1}{t^\delta}) F(\frac{x}{t^\delta})\] is valid, the corresponding entropy increases over time as \[S(t) = - \int_{\infty}^{+\infty}~{\rm d}x~p(x,t) ~\ln[p(x, t)].\] Using again the scaling condition for $p(x,t)$ one obtains \[S(t) = A + \delta \ln(t).\] With \[A = \int_{\infty}^{+\infty}{\rm d}y ~F(y)~ln[F(y)] = \mbox{constant},\] where $y = \frac{x}{t^\delta}$. This indicates that in the case of a diffusion process with a scaling probability distribution, its entropy $S(t)$ increases linearly with $\ln(t)$. The scaling exponent $\delta$ is being determined from the gradient of the fitted straight line in the linear-log plot of $S(t)$ over time $t$. For fractional Brownian motion the scaling exponent $\delta$ coincides with the Hurst exponent $H$. For random noise with finite variance, the probability distribution function $p(x, t)$ will converge, according to the central limit theorem, to a Gaussian distribution with $H = \delta = 0.5$ . If the Hurst exponent $H$ is different from the scaling exponent $\delta$, the scaling represents anomalous behavior. The diffusion process characterized by L\'{e}vy flights belongs to anomalous diffusion. In this specific case the scaling condition for $p(x,t)$ is still valid but the variance is not finite and the variance scaling exponent can not be defined. A second type of anomalous diffusion is due to L\'{e}vy walk based on a generalization of the central limit theorem. In this case of anomalous diffusion, the second moment is finite and the exponents H and delta obey the relation $\delta= 1/(3 - 2H)$.
In the Figure, the scaling exponent of $\delta = 0.52$ for the red straight line does indicate a deviation from Gauss. Also shown in green is the line for $\delta = 0.5$.\\

\noindent
{\bf Acknowledgment}\\

The authors would like to thank the Department of Science and Technology, Government of India, New Delhi, for the financial assistance for this work under project No. SR/S4/MS:287/05, and the Centre for Mathematical Sciences
for providing all facilities. Enlightening discussions with Professor A.M. Mathai, Director of the Centre for Mathematical Sciences, are gratefully acknowledged.

\end{document}